# Entropy production rate of nonequilibrium systems from the Fokker-Planck equation

Yu Haitao and Du Jiulin

*Department of Physics, School of Science, Tianjin University, Tianjin 300072, China*

**Abstract:** The entropy production rate of nonequilibrium systems is studied via the Fokker-Planck equation. This approach, based on the entropy production rate equation given by Schnakenberg from a master equation, requires information of the transition rate of the system under study. We obtain the transition rate from the conditional probability extracted from the Fokker-Planck equation and then derive a new and more operable expression for the entropy production rate. Examples are presented as applications of our approach.

## 1. Introduction

Nonequilibrium phenomena are ubiquitous in nature. A general description of nonequilibrium systems is therefore desirable. An available framework uses the notion of entropy production. It has been recognized that the distinguishing feature of a system out of thermodynamic equilibrium is the continuous production of entropy [1]. The variation of entropy per unit time can be divided into two parts: the exchange of entropy with the environment and the internal entropy production [2-4], that is

$$\frac{dS}{dt} = \frac{d_e S}{dt} + \frac{d_i S}{dt}, \tag{1}$$

where $S$ is the entropy of the system, $\frac{d_e S}{dt}$ is the flow of entropy per unit time between the environment and the system, $\frac{d_i S}{dt}$ is the internal entropy production rate. When a system is in a stationary state, $\frac{dS}{dt}$ vanishes and $\frac{d_i S}{dt} = -\frac{d_e S}{dt}$. If the stationary state is also an equilibrium state, then $\frac{d_i S}{dt} = 0$; and if it is a nonequilibrium state, $\frac{d_i S}{dt} > 0$.

Since the notion of entropy plays an important role in the development of statistic mechanics, the entropy production of a nonequilibrium system is regarded as a matter of primary importance. The interest on this quantity is not only focused on why the entropy increases as the nonequilibrium system evolves but also on how the entropy is produced. These two topics have been the subject of many theories.Perhaps, the most famous among them is the principle of minimum entropy production, introduced by Prigogine [4]. A much less known theorem, the principle of maximum entropy production was proposed by Jaynes [5] as a universal method for constructing the microscopic probability distributions of equilibrium and nonequilibrium statistical mechanics [6]. The fluctuation theorem, which is related to the probability distribution of the time-averaged irreversible entropy production, has been studied and developed in many aspects [7-12]. Even after such powerful theorems were proved, however, one can hardly say that the nonequilibrium problem has been solved. Much more research is needed.

This work describes a method to derive the entropy production rate for out of equilibrium systems. This new method follows a line analogous to Schnakenberg's; certain differences nonetheless distinguish the two methods. In section 2, we will first introduce a few basic notions, including the entropy produciton expression suggested by Schnakenberg on the basis of a master equation and a few elements of probability theory. In section 3, we combine the transition rate in a master equation with the transition probability density that describes a continuous stochastic variable, and in section 4, a new expression of the entropy production rate will be obtained. Examples will be presented in section 5, and finally our conclusion is summarized in section 6.

**2. Basic notions**

The common definition of a nonequilibrium Gibbs entropy is given by

$$S = -k_B \sum_n p_n(t) \ln p_n(t), \qquad (2)$$

where $k_B$ is Boltzmann constant and $p_n(t)$ is the probability of state $n$ at time $t$. For a system that can be described by the master equation [13],

$$\frac{\partial p_n(t)}{\partial t} = \sum_m \left[ p_m(t) w_{m,n}(t) - p_n(t) w_{n,m}(t) \right], \qquad (3)$$

where $w_{m,n}(t)$ is the transition rate from state $m$ to state $n$ at time $t$, Schnakenberg [14] suggested the following expression for the entropy production rate:

$$\begin{aligned}\frac{d_i S}{dt} &= \frac{k_B}{2} \sum_{m,n} [p_m(t) w_{m,n}(t) - p_n(t) w_{n,m}(t)] \ln \frac{p_m(t) w_{m,n}(t)}{p_n(t) w_{n,m}(t)} \\ &= k_B \sum_{m,n} p_m(t) w_{m,n}(t) \ln \frac{p_m(t) w_{m,n}(t)}{p_n(t) w_{n,m}(t)}. \end{aligned} \qquad (4)$$

This result was derived by constructing a fictitious system, a materially open homogeneous system containing chemical species and allowing chemical reactions between pairs of the components. Equation (4), which satisfies all the requirements imposed on the entropy production, has been considered by several authors [1,15-18].

One difficulty associated with the derivation of the entropy production from Eq.(4) is to obtain an expression for the transition rate $w_{m,n}(t)$. To circumvent this obstacle, we take advantage of the elementary probability theory. Consider a stochastic variable X, and let $p(x, t)$ be the probability density that X takes the value x at time t. We then have that

$$p(x,t+\tau) = \int dx' P(x,t+\tau | x',t) p(x',t), \tag{5}$$

where $P(x,t+\tau | x',t)$ is the conditional probability density for the variable X to take value x at time $t+\tau$ given that it had value $x'$ at time $t$.

If the process dynamics is Markovian, i.e., the memory of the stochastic variable is limited to its immediate past, then the conditional probability density $P(x,t+\tau | x',t)$ is the transition probability density [19]. We shall see that the two concepts $w_{m,n}(t)$ and $P(x,t+\tau | x',t)$ are nearly the same. They are different only because that the former is discrete, and the latter, continuous. The following derivation will show how $w_{m,n}(t)$ is linked to $P(x,t+\tau | x',t)$, after which a new expression of the entropy production rate for the continuous states will be derived. And to some extent, since $P(x,t+\tau | x',t)$ can be precisely determined, the new expression is more operable than the one following from Eq. (4).

## 3. From the transition probability density to the transition rate

If a nonequilibrium process described by a continuous stochastic variable X is Markovian, condition Eq. (5) is satisfied. One example is a Brownian particle in a viscous fluid, for which the time evolution of the probability density follows the Fokker-Planck equation. We want to describe the system states with a small quantity $\varepsilon$. If the stochastic variable X takes values between $(n-1/2)\varepsilon$ and $(n+1/2)\varepsilon$, we say that the system is in the *n*th state. And then that the probability of the system is in the *n*th state is the probability that the stochastic variable X takes the values between $(n-1/2)\varepsilon$ and $(n+1/2)\varepsilon$, that is

$$p_n(t) = \int_{n\varepsilon-\frac{\varepsilon}{2}}^{n\varepsilon+\frac{\varepsilon}{2}} d\xi \, p(x,t), \tag{6}$$

where $p(x,t)$ is the probability density.

As mentioned before, for a Markovian process, the conditional probability density $P(x,t+\tau\,|\,x',t)$ is the transition probability density. The difference between the transition probability density $P(x,t+\tau\,|\,x',t)$ and the transition rate $w_{m,n}(t)$ is that while $P(x,t+\tau\,|\,x',t)$ describes a continuous stochastic variable, $w_{m,n}(t)$ describes a discrete one. While the former is the transition during time $t \to t+\tau$, the latter is the transition per unit time.

Physically, $w_{m,n}(t)$ represents the probability of the transition from state $m$ to state $n$ per unit time. Therefore $w_{m,n}(t)\tau$ is the probability of the transition from state $m$ to state $n$ during the time interval $t \to t+\tau$. And $w_{m,n}(t)p_m(t)\tau$ is the joint probability that the system is in state $n$ at time $t+\tau$ and in state $m$ at time $t$. We use $p_{m,n}(t,\tau)$ to denote $p_{m,n}(t,\tau) = w_{m,n}(t)p_m(t)\tau$. The following equation is then a consequence of the elementary probability theory combined with Eqs.(5) and (6), that is

$$p_{m,n}(t,\tau) = \int_{n\varepsilon-\frac{\varepsilon}{2}}^{n\varepsilon+\frac{\varepsilon}{2}} dx \int_{m\varepsilon-\frac{\varepsilon}{2}}^{m\varepsilon+\frac{\varepsilon}{2}} dx'\, P(x,t+\tau\,|\,x',t)p(x',t), \qquad (7)$$

and so

$$w_{m,n}(t) = \frac{\int_{n\varepsilon-\frac{\varepsilon}{2}}^{n\varepsilon+\frac{\varepsilon}{2}} dx \int_{m\varepsilon-\frac{\varepsilon}{2}}^{m\varepsilon+\frac{\varepsilon}{2}} dx'\, P(x,t+\tau\,|\,x',t)p(x',t)}{\tau p_m(t)}. \qquad (8)$$

We now try to obtain simpler expressions for the discrete probability (6) and the transition rate (8) from the integral mean-value theorem. This theorem states that if $G:[a,b]\to \mathbf{R}$ (where $\mathbf{R}$ is the real domain) is a continuous function and $\varphi$ is an integrable function that does not change sign in the interval $(a,b)$, then value $x$ in $(a,b)$ can be found such that

$$\int_a^b G(t)\varphi(t)dt = G(x)\int_a^b \varphi(t)dt.$$

In particular, if $\varphi(t)=1$ for all $t$ in $[a,b]$, we have that

$$\int_a^b G(t)dt = G(x)(b-a).$$

The mean-value theorem reduces Eq. (7) to the relation,

$$p_n(t) = \int_{n\varepsilon-\frac{\varepsilon}{2}}^{n\varepsilon+\frac{\varepsilon}{2}} d\xi\, p(\xi,t) = \varepsilon p(\lambda,t), \qquad (9)$$

where $\lambda \in (n\varepsilon - \varepsilon/2, n\varepsilon + \varepsilon/2)$. When $\varepsilon$ is very small, $n\varepsilon$ can be approximately chosen as $\lambda$, so that

$$p_n(t) = \varepsilon p(n\varepsilon, t). \qquad (10)$$

With the same approach, we convert Eq. (8) to the form

$$w_{m,n}(t) = \frac{\varepsilon P(n\varepsilon, t+\tau \mid m\varepsilon, t) p_m(t)}{\tau p_m(t)} \\ = \frac{\varepsilon}{\tau} P(n\varepsilon, t+\tau \mid m\varepsilon, t). \qquad (11)$$

Equation (11) exactly shows the relation between the transition probability density and the transition rate.

## 4. Entropy production rate

Substitution of the right-hand sides of Eq.(10) and Eq.(11) for $p_n(t)$ and $w_{m,n}(t)$, respectively, in Eq. (4) yields the expression,

$$\frac{d_i S}{dt} = k_B \sum_{m,n} \frac{\varepsilon^2}{\tau} P(n\varepsilon, t+\tau \mid m\varepsilon, t) p(m\varepsilon, t) \ln \frac{P(n\varepsilon, t+\tau \mid m\varepsilon, t) p(m\varepsilon, t)}{P(m\varepsilon, t+\tau \mid n\varepsilon, t) p(n\varepsilon, t)}. \qquad (12)$$

With $\varepsilon \to 0$, the sum in the above equation turns into an integration, and we have that

$$\frac{d_i S}{dt} = k_B \int dx \int dx' \frac{1}{\tau} P(x, t+\tau \mid x', t) p(x', t) \ln \frac{P(x, t+\tau \mid x', t) p(x', t)}{P(x', t+\tau \mid x, t) p(x, t)}, \qquad (13)$$

which is a new expression for the entropy production rate for a continuous stochastic variable. To proceed and derive the entropy production rate from Eq.(13), we need an exact expression for $P(x, t+\tau \mid x', t)$. Fortunately, for the one-variable Fokker-Planck equation [20],

$$\frac{\partial p(x,t)}{\partial t} = -\frac{\partial}{\partial x}\left(D^{(1)}(x,t) p(x,t)\right) + \frac{\partial^2}{\partial x^2}\left(D^{(2)}(x,t) p(x,t)\right), \qquad (14)$$

where $D^{(1)}(x,t)$ and $D^{(1)}(x,t)$ are the drift coefficient and the diffusion coefficient respectively, the conditional probability density $P(x, t+\tau \mid x', t)$ can be expressed as

$$P(x,t+\tau|x',t) = \left[1 - \tau D^{(1)}(x',t)\frac{\partial}{\partial x} + \tau D^{(2)}(x',t)\frac{\partial^2}{\partial x^2} + O(\tau^2)\right]\delta(x-x'). \tag{15}$$

The omission of the high-order small quantities, $O(\tau^2)$, and the use of the Fourier integral representation of $\delta$ function then yield the result,

$$P(x,t+\tau|x',t) = \frac{1}{2\sqrt{\pi D^{(2)}(x',t)\tau}}\exp\left(-\frac{[x-x'-D^{(1)}(x',t)\tau]^2}{4D^{(2)}(x',t)\tau}\right). \tag{16}$$

There are two ways to obtain an exact formula of the entropy production for a nonequilibrium process. One is to insert Eq. (16) into Eq. (13),

$$\frac{d_i S}{dt} = k_B \int dx \int dx' \frac{p(x',t)e^{-\frac{[x-x'-a(x')\tau]^2}{4b(x')\tau}}}{2\tau\sqrt{\pi b(x')\tau}} \ln \frac{\sqrt{b(x)}p(x',t)e^{-\frac{[x-x'-a(x')\tau]^2}{4b(x')\tau}}}{\sqrt{b(x')}p(x,t)e^{-\frac{[x'-x-a(x)\tau]^2}{4b(x)\tau}}}, \tag{17}$$

with $a(x) = D^{(1)}(x,t)$ and $b(x) = D^{(2)}(x,t)$.

The other is to substitute the right-hand side of Eq.(15) for $P(x,t+\tau|x',t)$ before the logarithm on the right-hand side of Eq.(13) and carry out the integration over $x'$ with the help of $\delta$ function to obtain the result,

$$\begin{aligned}\frac{d_i S}{dt} = &k_B \int dx D^{(1)}(x,t)p(x,t)\left[\frac{\partial}{\partial x'}\ln\frac{P(x',t+\tau|x,t)p(x,t)}{P(x,t+\tau|x',t)p(x',t)}\right]_{x'=x} \\ &+ k_B \int dx D^{(2)}(x,t)p(x,t)\left[\frac{\partial^2}{\partial x'^2}\ln\frac{P(x',t+\tau|x,t)p(x,t)}{P(x,t+\tau|x',t)p(x',t)}\right]_{x'=x}.\end{aligned} \tag{18}$$

Insertion of Eq.(16) into Eq.(18) yields an expression that maybe different from Eq. (17), but is more manageable. Either way yields an exact expression, distribution-dependent for the entropy production.

Notwithstanding the exact Eq.(17), a difficulty rises. Eq.(14) is usually used to describe a system whose states are continuous, such as a Brownian motion. Nevertheless we have used the probability (4), with discrete states described by a master equation. When we try to calculate the entropy production rate with Eq.(4), should we use the Fokker-Planck equation once again? We see that all the derivations above and the ones below, rely on Eq.(5). From Eq.(5), one can derive the master equation [19] and also the Fokker-Planck equation [20], provided only that the Markov condition is satisfied. We see that the master equation and the Fokker-Planck equation are two different phenomenological descriptions for one system. As Eq.(17) shows, these descriptions denend on the parameter $\tau$, which describes the shortest transition time, which sets the time scale of the target system.

## 5. Examples of the Smoluchowski equation and Ornstein-Uhlenbeck process

An important example of one-variable Fokker-Planck equations is the Smoluchowski equation, which describes the probability distribution of a Brownian particle moves in a strong friction medium [21]. In this case, the coordinate undergoes a creeping motion leading to Smoluchowski equation,

$$\frac{\partial p(x,t)}{\partial t} = \frac{1}{m\gamma}\left[-\frac{\partial}{\partial x}F(x) + k_B T \frac{\partial^2}{\partial x^2}\right]p(x,t), \tag{19}$$

where $m$ is the mass of the Brownian particle, $\gamma$ is the friction coefficient and $F(x)$ is the force acting on the particle. The comparison between Eq.(19) and Eq.(14) yields

$$D^{(1)}(x,t) = \frac{F(x)}{m\gamma}; \quad D^{(2)}(x,t) = \frac{k_B T}{m\gamma}. \tag{20}$$

In inserting Eq.(20) into Eq. (16), we then find that

$$P(x,t+\tau\,|\,x',t) = \frac{1}{2\sqrt{\pi b\tau}}\exp\left(-\frac{[x-x'-a(x')\tau]^2}{4b\tau}\right), \tag{21}$$

with $a(x) = F(x)/m\gamma$ and $b = k_B T/m\gamma$.

Insertion of Eq. (21) into Eq. (13) yields the equality

$$\frac{d_i S}{dt} = k_B \int dx \int dx' \frac{p(x',t)e^{-\frac{[x-x'-a(x')\tau]^2}{4b\tau}}}{2\tau\sqrt{\pi b\tau}} \ln \frac{p(x',t)e^{-\frac{[x-x'-a(x')\tau]^2}{4b\tau}}}{p(x,t)e^{-\frac{[x'-x-a(x)\tau]^2}{4b\tau}}}. \tag{22}$$

Alternatively, the insertion of Eq. (21) into Eq. (18) followed by the integration by parts and the limit $\tau \to 0$ shows that

$$\frac{d_i S}{dt} = k_B \int \frac{[J(x,t)]^2}{bp(x,t)}dx, \tag{23}$$

where

$$J(x,t) = a(x)p(x,t) - b\frac{\partial}{\partial x}p(x,t) \tag{24}$$

is the probability current in Eq. (19).

Eq.(24) is equivalent to the result derived by Tomé in the multi-particles case [22]. And as the by-product, the entropy flux term, which is the negative of the entropy production rate, is also the same. Our approach to calculate the entropy production rate is equivalent to Tomé's, but the equivalence is limited to constant diffusion coefficients. When the diffusion coefficient is a function of $x$, Eq. (23) will be no longer valid, and even the probability current will no longer be given by Eq. (24).

To derive and express the entropy production rate from Eq.(22), we need to know $F(x)$. For the simple case, $F(x) = 0$, Eq. (19) reduces to the diffusion equation,

$$\frac{\partial p(x,t)}{\partial t} = D \frac{\partial^2 p(x,t)}{\partial x^2}. \tag{25}$$

Eq.(25) describe a Wiener process. The solution is

$$p(x,t) = \frac{1}{2\sqrt{\pi D t}} \exp\left(-\frac{x^2}{4Dt}\right), \tag{26}$$

with the initial condition $p(x,0) = \delta(x)$.

The transition probability is simply

$$P(x, t+\tau \mid x', t) = \frac{1}{2\sqrt{\pi D \tau}} \exp\left(-\frac{(x-x')^2}{4D\tau}\right). \tag{27}$$

When we insert Eqs.(26) and (27) into Eq. (13), the following expression for the entropy production rate of the diffusion process is easy obtained,

$$\frac{d_i S}{dt} = \frac{k_B}{2t}. \tag{28}$$

The parameter $\tau$ has disappeared in Eq.(28). It is clear that for a Wiener process, the entropy production rate will reduce to zero when $t \to \infty$, as the system approaches equilibrium.

Another important example is the Ornstein-Uhlenbeck process, with

$$D^{(1)}(x) = -\gamma x; \quad D^{(2)}(x) = D = \text{const.} \tag{29}$$

The probability distribution equation for this process

$$\frac{\partial p(x,t)}{\partial t} = \gamma \frac{\partial}{\partial x}(xp(x,t)) + D \frac{\partial^2 p(x,t)}{\partial x^2}. \tag{30}$$

has the following stationary solution $p(x)$, which satisfies $\partial p / \partial t = 0$ for $\gamma > 0$,

$$p(x) = \sqrt{\frac{\gamma}{2\pi D}} \exp\left(-\frac{\gamma x^2}{2D}\right), \tag{31}$$

and the corresponding equation reduces to

$$\gamma \frac{\partial}{\partial x}(xp(x)) + D \frac{\partial^2 p(x)}{\partial x^2} = 0 \tag{32}$$

with the boundary conditions $p(x) \to 0$ when $x \to \pm\infty$.

With the initial condition $P(x,t|x',t) = \delta(x-x')$ and $p(x) = \int_{-\infty}^{\infty} P(x,t+\tau|x',t)p(x')dx'$, Eq.(30) shows that the transitions probability density $P(x,t+\tau|x',t)$ [20] is

$$P(x,t+\tau|x',t) = \sqrt{\frac{\gamma}{2\pi D(1-e^{-2\gamma\tau})}} \exp\left[-\frac{\gamma(x-e^{-\gamma\tau}x')^2}{2D(1-e^{-2\gamma\tau})}\right]. \quad (33)$$

We now insert Eqs.(31) and (33) into Eq.(13). The entropy production rate $\frac{d_i S}{dt} = 0$ then results, which shows that Eq.(31) is an equlibrium state of the Ornstein-Uhlenbeck process.

## 6. Conclusion

We have presented an approach to calculate the entropy production rate based on Eq.(4), given by Schnakenberg. We have used the conditional probability density easily obtained from the Fokker-Planck to derive the transition rate in the master equation description and then obtain the new expression (17) of the entropy production rate.

As compared with the Schnakenberg's expression Eq.(4), our entropy production rate Eq.(17) is more operable. The entropy production rate can be calculated exactly if the drift and diffusion coefficients are known. By contrast, the transition rate in Eq.(4) is usually difficult to calculate, unless phenomenological methods or conjectures are adopted.

**Acknowledgments**

This work is supported by the National Natural Science Foundation of China under Grant No. 11175128.